\documentstyle[11pt,moriond,epsfig]{article}

\def\Journal#1#2#3#4{{#1} {\bf #2}, #3 (#4)}

\def\NPB{{\em Nucl. Phys.} B}
\def\PLB{{\em Phys. Lett.}  B}

\def\PRD{{\em Phys. Rev.} D}

\def\be{\begin{equation}}
\def\ee{\end{equation}}
\def\bea{\begin{eqnarray}}
\def\eea{\end{eqnarray}}

\begin{document}

\rightline{KUCP-0181}
\rightline{hep-th/0105080}

\vspace*{2.5cm}

\title{GRAVITATIONAL STABILITY AND SCREENING EFFECT FROM EXTRA TIMELIKE 
DIMENSIONS
\footnote{Invited talk presented at the XXXVIth 
Rencontres de Moriond ``QCD and High Energy Hadronic 
Interactions", Les Arcs, Savoie, France, March 17-24, 2001.}
}

\author{ SATOSHI MATSUDA 
\footnote{E-mail address:
matsuda@phys.h.kyoto-u.ac.jp}
}

\address{Department of Fundamental Sciences, FIHS, Kyoto University\\
Yoshida, Sakyo-ku, Kyoto 606-8501, Japan}

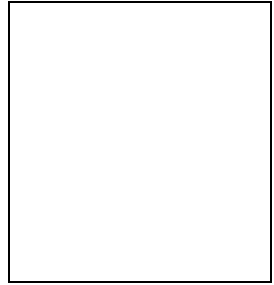
\maketitle\abstracts{
We discuss extra timelike dimensions and their effects on the 
gravitational stability of spherical massive bodies. 
Here we specifically report our results for the case of one extra timelike 
dimension where we have made analytically rigorous investigations on the 
tachyonic graviton exchange due to the infinite tower of the 
Kaluza-Klein mode. 
With the scale $L$ of the extra timelike dimension we find that 
some spherical bodies of radius $R$ can be stable at critical radii 
$R=2\pi Lp$ for some positive integer $p$. We also obtain the generic 
property of massive bodies that for the short distance range 
$0<R\leq \pi L$ the gravitational force due to the ordinary massless 
graviton exchange is screened by the Kaluza-Klein mode exchange of 
tachyonic gravitons.}

\section{Physics of Extra Timelike Dimensions}

\subsection{Possibility for extra times}

There is no a priori reason why extra times cannot exist. 
A possibility of the existence of $D$ extra timelike dimensions is 
an interesting subject in its own right and has also been 
discussed in various contexts in the past:
the subject has been studied in cosmological constant problem by 
Aref'eva {\it et al}~\cite{aref}, in string theories 
by Vafa~\cite{vaf} {\it et al}, 
in brane theories by Chaichian {\it et al}~\cite{chai}, 
in supergravity by Popov~\cite{pop} {\it et al}, 
and in many other topics such as the ones  
by Sakharov~\cite{sak}, Chamblin {\it et al}~\cite{ce}.

\subsection{Experimental Viability and Signatures}
  
Phenomenology and observable physical effects of extra timelike 
dimensions have been discussed by Yndur$\acute {\rm a}$in~\cite{yn} 
and by Dvali {\it et al}~\cite{dva} in a scenario where 
the Standard Model particles 
are localized in ``our time", whereas gravity can propagate in 
all time directions.

The compactification of extra timelike space gives rise to 
the Kaluza-Klein mode of tachyons and causes the violation of 
causality and conserved probability. The gravitational potential 
turns out to have an imaginary part which makes massive bodies 
unstable. But the observable effect is not unacceptable 
experimentally if the scale $L$ of the extra dimensions is 
bounded below a sufficiently small size~\cite{yn} 
which is nearly as close as to the Planck scale. 

One may naturally ask what then are other noticeable signatures of 
extra timelike dimensions. In this report we aim at definite 
rigorous study of this question being based on an analytic 
ground.

\section{Gravitational Stability and Screening Effect}
\subsection{Gravitational Potential}

We consider one extra timelike dimension with $D=1$ and 
compactify it on a circle of radius $L$. 
We shall look into the problem of the Kaluza-Klein tachyonic mode 
in detail based on an analytically precise treatment, thus 
obtaining the rigorous answer to the problem of instability of 
massive bodies~\cite{ms1}. 

Let gravitons propagate in the extra dimension; 
then we obtain tachyonic gravitons of the Kaluza-Klein mode. 
Their propagators are:
\begin{equation}
-i{1\over k_0^2-\mbox{\boldmath $k$}^2+{n^2\over L^2}+i\epsilon},
\quad n=integers
\label{eq:prop}
\end{equation}
up to a spin tensor factor. Then the gravitational potential 
between two unit mass points at distance $d$ is given by 
\begin{equation}
V(d)=-G_{\rm N}{1\over d}-\sum_{n=-\infty,n\not=1}^{+\infty}G_{\rm N}
{1\over d}e^{+i{|n|\over L}d} \sim {1\over d^{1+1}}
\quad as\quad d\rightarrow 0
\label{eq:gpot}
\end{equation}
in the nonrelativistic tree-level approximation, where $G_{\rm N}$ 
is the Newton constant.

We point out here that the complex gravitational potential used by 
Yndur$\acute {\rm a}$in~\cite{yn} and also 
by Dvali {\it et al}~\cite{dva} has a wrong sign 
in the phase factor, and that the correct sign of the phase together 
with the correct overall sign of the potential is crucially important 
in discussing the stability of matter from the aspect of its 
{\em vanishing} or {\em explosion}.

\subsection{Gravitational Self-Energy}
The gravitational self-energy of a spherical body of radius $R$ 
with mass density $\rho(r)$ is calculated as 
\begin{equation}
E_n(R)=8i\pi^2G_{\rm N}L\int_0^R dr \int_0^r d\ell\  
\rho(r)\rho(\ell){r\ell\over |n|} 
\Bigl[e^{i{|n|\over L}(r+\ell)}-e^{i{|n|\over L}(r-\ell)}   \Bigr],   
\label{eq:selfn}
\end{equation}
which gives for $n\rightarrow 0$
\begin{equation}
E_0=-16\pi^2G_{\rm N}\int_0^R dr \int_0^r d\ell\ \rho(r)\rho(\ell)r\ell^2.
\label{eq:selfzero}
\end{equation}
The total gravitational self-energy is then:
\begin{equation}
E(R)=E_0+\sum_{n=-\infty,n\not=0}^{+\infty}E_n=E_0+
2\sum_{n=1}^\infty E_n.
\label{eq:selftot}
\end{equation}

\subsection{Gravitational Stability and Screening Effect}

Let us choose the mass density $\rho(r)=D/r$. Then we get 
the imaginary part of the self-energy as  
\begin{equation}
\Im E(R=2\pi Lk+c)=-16\pi^2G_{\rm N}D^2L\int_0^c dr \int_0^r d\ell\ 
\ln{\Bigl|\sin{r+\ell\over 2L}\Bigr|\over \sin{r-\ell\over 2L}}, \quad
0\le c<2\pi L
\label{eq:imself}
\end{equation} 
for non-negative integers $k\ge 0$, while we obtain the real part of 
the self-energy as 
\begin{equation}
\Re E(R=2\pi Lk+c)=-32\pi^3G_{\rm N}D^2L
\cases{
 f(k, c)          & if $0\le c <\pi L$     \cr
 f(k+1, c-2\pi L) & if $\pi L\le c<2\pi L$ \cr
        }
\label{eq:reself}
\end{equation}
where we have defined the function $f(k,c)$ by 
\begin{equation}
f(k,c)\equiv k\big[3c^2+6\pi Lkc+(4k^2-1)\pi^2L^2\big].
\end{equation}

Obviously $\Im E(R)$ is periodic in R and vanishes at radii 
$R=2\pi Lk$ with $c=0$ (see Fig. 1 of Matsuda and Seki~\cite{ms1}). 
This implies that the spherical massive body becomes stable 
at the critical radii. 

We also note that $Re E(R)$ identically vanishes for the very short 
range $0\le R\le \pi L$ and turns into negative values for the longer 
range $R>\pi L$ (see Fig. 2 of Matsuda and Seki~\cite{ms1}). 
This suggests that for the above short range the ordinary 
gravitational potential due to the exchange of massless graviton is 
completely ``screened" by the tachyonic graviton exchange of the infinite 
tower of the Kaluza-Klein mode.   
 
\subsection{Generic Features of Gravitational Stability and 
Screening Effect}

So far we have chosen one particular type of mass density. 
One could construct an {\em onionlike hybrid model} for the mass 
density:
\begin{equation}
\rho_{\rm H}=\cases{ 
{D\over r} & for max$[0,(2km-1)\pi L]<r<(2km+1)\pi L$  \cr
{bD\over r} & for $(2km+1)\pi L<r<\big(2(k+1)m-1\big)\pi L$, \cr
     }\quad k=0,1,2,\dots
\label{eq:hybrid}
\end{equation}  
where $m$ is a fixed positive integer and $b$ is a positive constant. 
One easily sees that, as $b\rightarrow 1$ or when $m=1$, 
$\rho_{\rm H}(r)\rightarrow \rho(r)=D/r$. 
By varying $b$ and $m$, we obtain a variety of 
{\em onionlike hybrid models} 
with the common generic feature of gravitational stability which 
shows up in each model at critical radius $R=2\pi Lp$ with a 
corresponding positive integer $p=km$:
\begin{equation}
\Im E(R=2p\pi L)=\Im E(R=2km\pi L)=0
\end{equation}

The gravitational screening effect turns out to be the generic 
feature of the Kaluza-Klein mode with an extra timelike dimension. 
This can be proved rigorously by the use of the summation formula:
\begin{equation}
i\sum_{n=1}^\infty {1\over n}\Bigl[e^{i{n\over L}(r+\ell)}- 
e^{i{n\over L}(r-\ell)}\Bigr]={1\over L}-
i\ln{\sin{r+\ell\over 2L}\over \sin{r-\ell\over 2L}},\quad
0<r-\ell\le r+\ell<2\pi L.
\label{eq:sumf}
\end{equation}
For $0<R\le \pi L$ and any choice of $\rho(r)$ we obtain 
from Eq.~\ref{eq:selfn}, Eq.~\ref{eq:selfzero} and Eq.~\ref{eq:selftot}:
\begin{equation}
E(R)=-i16\pi^2G_{\rm N}L\int_0^R dr\int_0^r d\ell\ \rho(r)\rho(\ell)
r\ell \ln{\sin{r+\ell\over 2L}\over \sin{r-\ell\over 2L}},\quad
0<R\le \pi L
\end{equation}
which proves that the real part of the self-energy vanishes identically 
for the short range $0<R\le \pi L$ for any spherical mass density.
We can therefore conclude generically that in the region 
$0<R\le \pi L$ the gravitational force due to the ordinary massless 
graviton exchange is ``screened" by the effect of the tachyonic 
graviton exchange of the Kaluza-Klein mode.

\section{Conclusions}
\subsection{Comments}

We note that, in this paper, by ``gravitational stability" we mean 
a stability of the self-energy state of a spherical massive body 
in pure Newtonian gravity, not the stability under metric 
perturbation usually referred to in Einstein gravity.

We also note that our results are classical, and it is not clear 
how the quantum loop corrections would modify the picture presented 
above. It is also not clear whether for the very short distance 
under our consideration a four-dimensional effective theory is 
well adapted to describe the physics for spherical massive bodies 
investigated here.

But with no concrete quantum gravity theory available for such small 
distances, we have performed our study by assuming that a conventional 
physical reasoning of our four-dimensional world is valid.

Though we have considered one extra timelike dimension, there is 
no reason why no additional timelike dimensions could exist. 
The extended case of $D (\ge 1)$ extra times has been studied 
in detail~\cite{ms2}.

\subsection{Summary}  

Extra timelike as well as spacelike dimensions have rich physics 
structure for solving the standing problems in current particle 
theory such as the hierarchy problem and the cosmological constant 
problem. 

We have made rigorous analytic studies on extra timelike dimensions 
and have shown on an analytic ground that some spherical bodies 
can be gravitationally stable at critical radii $R=2\pi L p$ 
for some positive integer $p$.

We have also proved the generic property of massive bodies that 
for the range $0<R\le \pi L$ the gravitational force due to the 
ordinary massless graviton exchange is screened by the Kaluza-Klein 
mode exchange of tachyonic gravitons.

\section*{Acknowledgments}
This work is supported in part by the Grant-in-Aid for Scientific 
Research on Priority Area 707 ``Supersymmetry and Unified Theory of 
Elementary Particles", Japan Ministry of Education, Culture, Sports, 
Science and Technology, and also 
by the Grant-in-Aid for Scientific Research (C)(2)-10640260.


\section*{References}
\begin{thebibliography}{99}

\bibitem{aref}I.Ya. Aref'eva {\it et al}., \Journal{\PLB}{177}{357}{1986}.

\bibitem{vaf}C. Vafa, \Journal{\NPB}{496}{403}{1996}.

\bibitem{chai}M. Chaichian and A.B. Kobakhidze, \Journal{\PLB}{488}
{117}{2000}.

\bibitem{pop}A. Popov, \Journal{\PLB}{259}{256}{1991}.

\bibitem{sak}A.D. Sakharov, {\em Sov. Phys. JETP} {\bf 60}, 214 (1984).

\bibitem{ce}A. Chamblin and R. Emparan, \Journal{\PRD}{55}{754}{1997}.

\bibitem{yn}F.J. Yndur{$\acute{\rm a}$}in, \Journal{\PLB}{256}{15}{1991}.

\bibitem{dva}G. Dvali, G. Gabadadze and G. Senjanovi{$\acute {\rm c}$}, 
hep-ph/0007290.

\bibitem{ms1}S. Matsuda and S. Seki, \Journal{\PRD}{63}{065014}{2001}.

\bibitem{ms2}S. Matsuda and S. Seki, \Journal{\NPB}{599}{119}{2001}.

\end{thebibliography}
\end{document}